# Exploring Aspects of Polyglot High-Performance Virtual Machine GraalVM

M. Šipek, B. Mihaljević and A. Radovan
Rochester Institute of Technology Croatia, Zagreb, Croatia
matija.sipek@mail.rit.edu, branko.mihaljevic@croatia.rit.edu, aleksander.radovan@croatia.rit.edu

*Abstract* - Contemporary software often becomes vastly complex, and we are required to use a variety of technologies and different programming languages for its development. As interoperability between programming languages could cause high overhead resulting in a performance loss, it is important to examine how a current polyglot virtual machine with a compiler written in a high-level object-oriented language deals with it. OpenJDK's Project Metropolis presented the GraalVM, an open-source, high-performance polyglot virtual machine, mostly written in Java. This paper presents GraalVM's architecture and its features; furthermore, examining how it resolves common interoperability and performance problems. GraalVM makes software ecosystem productive when combining various programming languages, for example, Java, JavaScript, C/C++, Python, Ruby, R, and others. The vital part of GraalVM is the Graal compiler written in Java, which allows developers to maintain and optimize code faster, simpler, and more efficient, in comparison to traditional compilers in C/C++ languages. Graal can be used as a just-in-time (JIT) or as static, ahead-of-time (AOT) compiler. Graal is an aggressively optimizing compiler implementing common compiler optimizations, with emphasis on outstanding inlining and escape analysis algorithms. This paper compares Graal with some of the best-specialized competitors, and presents our results tested within an academic environment.

## I. INTRODUCTION

Process Virtual Machines (or Application Virtual Machines), designed to run as regular applications within an operating system (OS), provide a platform-independent programming environment and allow any compatible program execution, regardless of the platform, the OS, and the underlying hardware [1]. As high-level object-oriented programming (OOP) languages became more popular, new Virtual Machines (VMs) were implemented with an interpreter in order to deal with high-level abstraction from the details of the platform and OS. Many OOP implementations appeared using Process VMs; however, the popularity was enhanced with the rise of Java Virtual Machine (JVM) and .NET Framework's Common Language Runtime (CLR), which had the ability to automate important areas of program execution, such as memory management and garbage collection.

Nowadays, many different scripting and OOP languages are being used to create applications which run within VMs to solve various specific computing problems in a way to balance between simplicity and high performance. Moreover, in some cases we need to combine different languages to achieve an optimal solution for a given problem. We assembled a list of the most popular languages (Table I) which presents that no language overrules the others in popularity, and within the past years the difference between them decreased even more.

Cross-Language Interoperability [2] enables reuse of existing libraries in other languages but often requires additional steps, i.e., paying a high serialization cost. In that case program data needs to be marshaled and converted into an adequate format and communicated using some type of an interface. Afterward, it needs to be deserialized to use it in other languages, which comes with a performance cost.

Programming languages that are primarily used for high-performance purposes are being additionally specifically optimized to run in high-performance mode. However, that is most often not the case with the most popular general-purpose programming languages, which without such additions do not perform well in high-performance cases. Therefore, it is sometimes necessary to combine different languages in a cross-language interoperability manner to achieve the optimal results. When combining various languages, it is desired to use the same set of tools for configuration, debugging, and profiling to achieve interoperability, but at this moment that is rarely the case.

TABLE I. THE MOST POPULAR PROGRAMMING LANGUAGES[1][2] IN JANUARY 2019 [3][4]

| Language | TIOBE Index | PYPL Index | IEEE Spectrum Ranking | Google Trends Average |
|---|---|---|---|---|
| **Java** | 16.904% | 21.42% | 97.5 | 42 |
| **C** | 13.337% | 6.31%* | 96.7 | 11 |
| **Python** | 8.294% | 25.95% | 100 | 12 |
| **C++** | 8.158% | 6.31%[a] | 99.7 | 11 |
| **VB .NET** | 6.459% | 1.13% | 45.1 | 11 |
| **JavaScript** | 3.302% | 8.26% | 82.6 | 20 |
| **C#** | 3.284% | 7.62% | 89.4 | 12[b] |
| **PHP** | 2.680% | 7.37% | 84.9 | 16 |
| **Objective-C** | 1.781% | 3.15% | 50.5 | 1 |
| **R** | 1.331% | 4.04% | 82.9 | n/a |

a. A shared percentage for both C and C++ programming languages
b. Search term as a word, not as a programming language (not available)

---

[1] TIOBE Index, https://www.tiobe.com/tiobe-index/

[2] Google Trends, https://trends.google.com/

For the purpose of this research, we tested the performance of the current JVM and associated compilers, and compared it with GraalVM[3], a contemporary virtual machine with polyglot programming capabilities. In addition, we added a set of tests to determine the polyglot abilities of GraalVM, and presented architecture and different aspects of GraalVM, as well as preliminary results of conducted tests on induced features.

*A. About GraalVM*

One of the projects to circumvent the problems mentioned above and support wider and better cross-language interoperability on the JVM is OpenJDK's Graal Project[4], which evolved out of the Maxine VM project [5], featured as the next-generation modular architecture platform written in Java, compatible with modern Java IDEs and standard JDK [6].

As a part of the project, GraalVM was introduced, presenting a high-performance universal polyglot virtual machine that is able to run different programming languages with a minimal loss in performance. GraalVM enables usage of other programming languages with almost zero overhead interoperability when using external libraries.

GraalVM works efficiently with polyglot applications, thus allowing developers to choose the most appropriate language to deal with the problem without having to compromise performance [7]. GraalVM is designed to execute programs within the JVM using a Just-In-Time (JIT) compiler, but also using Ahead-Of-Time (AOT) compilation for creating native images or embedded into both managed and native applications. Polyglot capabilities can be extended by embedding GraalVM into different runtime platforms, for example, relational databases which can directly use programming languages.

Languages that GraalVM currently supports include:

- JVM-based languages – Java, Scala, Kotlin, Groovy, Clojure
- Interpreted languages – JavaScript, Python, R, Ruby
- Native languages – C, C++, Rust, Swift, Fortran

GraalVM is an open source project, mostly written in Java and supported by Oracle and its affiliates. The most important components of the GraalVM project are the Just-In-Time (JIT) compiler, also named Graal, and a framework used for implementation of other programming languages named Truffle.

## II. GRAALVM ARCHITECTURE

*A. Graal Project background*

Graal compiler from the OpenJDK Graal project started in 2012, as a part of the Maxine VM project, with a goal to create an advanced compiler in high-level OOP language, in this case, Java [8]. A starting concept was to design a dynamic compiler that produces high-quality code, simplifies development, and lowers the abstraction when using low-level languages, without compromising on compile times and memory usage. Future development was reassigned as Project Metropolis[5], which objective is to present advanced Java-On-Java implementation techniques for the HotSpot VM replacement.

*B. The Java HotSpot™ Virtual Machine*

The Java HotSpot VM is an implementation of the JVM, originally developed by Sun Microsystems, later acquired by Oracle, which became default JVM in Java 1.3. It is a high-performance VM that connects to the lowest stack of GraalVM architecture (Fig. 1); however, by itself, it can be used only for languages that target the JVM itself. This VM was designed in a way to seek frequently executed methods or "hot spots" in the code for JIT compiling optimization.

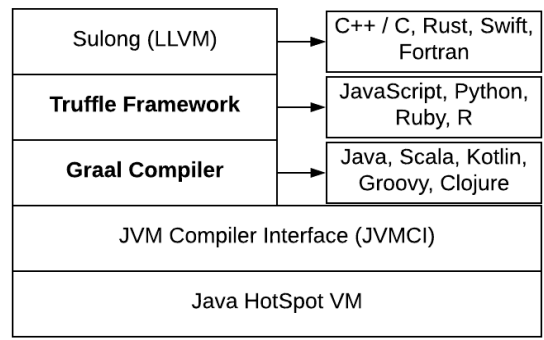

Figure 1. GraalVM Architecture

The HotSpot VM contains two JIT compilers which both compile bytecode to optimized machine code [9]:

- The server compiler also called *opto* or C2
- The client compiler also called C1

The server compiler creates highly optimized compiled code, mostly intended for long-running server applications since it uses a bottom-up tree rewriting component and a program dependence graph. The client compiler is based on a control flow graph so it can quickly generate compiled code and is mostly a better fit for desktop applications. C1 is tuned for quick loading and makes use of interpretation, as, on the other hand, C2 has slower loading with more precise and better optimizations which result in higher performance [9]. In addition, HotSpot VM uses two different interpreters, the C++ interpreter and the template interpreter, and contains memory management that performs garbage collection.

*C. JVM Compiler Interface*

The JVM Compiler Interface [6] (JVMCI) allows to implement a custom optimizing JIT compiler written in Java, which can be used in JVM as a dynamic compiler. In GraalVM architecture it is placed above the Hotspot VM (Fig. 1) and connected to the Graal compiler. JVMCI provides the possibility to use a compiler written in Java,

---

[3] GraalVM, https://www.graalvm.org/
[4] Graal Project, http://openjdk.java.net/projects/graal/
[5] Project Metropolis, https://openjdk.java.net/projects/metropolis/
[6] JEP 243: Java-Level JVM Compiler Interface, http://openjdk.java.net/jeps/243

which is easier to manage and improve than existing compilers written in C or C++.

*D. Graal, the JIT compiler*

An important part of GraalVM is Graal, a high-performance compiler, mainly used for JIT compilation, but also static, AOT compiler. As GraalVM has a modular architecture, a compiler is separated from VM, and Graal replaces C1 and C2 compilers used in HotSpot. This architecture provides the ability to reuse all VM components that don't interfere with compilation process, such as an interpreter, GC, Java Native Interface (JNI), class loading, and others [9].

Since compilation is a complex process, within GraalVM it is separated into two parts (Fig. 2), first of which is a source-specific, where most high-level optimizations are performed. This part includes a front end and platform-independent optimization phase named Graal Intermediate Representation (Graal IR). The second, target-specific part represented as Back End, is responsible for the translation of high-level Graal IR into low-level IR (LIR).

The front end is in control for turning Source representation as bytecode into Graal IR, using Truffle framework or Graphbuilder. Graphbuilder parses an array of bytes as JVM bytecode into a Graal graph and combines profiling feedback from interpreter [9]. On the other hand, if we are using Truffle framework [10], Graal recognizes this and aggressively optimizes its interpreter, allowing us to develop runtime for different programming languages.

After bytecode is interpreted into high–level intermediate representation Graal IR, the optimization process can begin. This is the generic part of compilation where all, platform-independent optimizations are performed. Graal IR is a complex hybrid structure which controls flow and data dependencies, which consists of two directed acyclic graphs and splits it optimization process into three tiers: High Tier, Mid Tier, and Low Tier. Tiers are deconstructed into independent phases, and at the end of each tier, a preparation named lowering phase deconstructs operations, preparing them for the next tier.

High Tier focuses on high-level optimizations, where common compiler optimizations are used, but the emphasis is on Graal's outstanding method inlining and partial escape analysis. Graal performs Late Inlining by parsing methods independently and combining them in a later compilation process. Thus, it is possible to cache the results of bytecode parsing to make better inlining decisions. Partial escape analysis verifies if an allocated object can be used outside of allocating thread or method. The object is being allocated on current thread's stack, or directly pushed to registers in order to avoid allocation [11]. Mid Tier mainly deals with memory optimizations, and Low Tier does the final clean up, small optimizations, and graph preparation for low-level representation conversion [9].

Finally, Back End, which deals with the translation of Graal IR into LIR, performs register allocation and forwards byte-code instructions to the processor.

---

[7] The Truffle Language Implementation Framework, https://github.com/oracle/graal/tree/master/truffle

*E. The Truffle Language Implementation Framework*

Developing an interpreter for a programming language should be more efficient than creating a high-performance compiler from scratch. Truffle [7] is an open-source framework for creating self-optimizing interpreters, for implementation of programming languages. Truffle uses an abstract syntax tree (AST) concept to implement a specific version of an interpreter by modifying it during the interpretation process to incorporate type feedback. This is achieved by using information of a guest programming language and a given program [12].

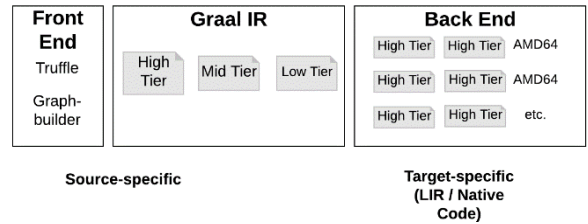

Figure 2. Compilation Process within GraalVM

As data is running through the program, several techniques could be used for optimization by specialization, such as inlining and constant folding. Graal incorporates logic for different types of optimizations, which are bundled in the Partial Evaluation technique; thus, Truffle uses Graal to create a Partial Evaluator.

The optimization process (Fig. 3) presents AST interpreter as the starting point of Truffle, by itself a technique that isn't sufficiently optimized. In the beginning, AST interpreter is populated by uninitialized nodes ("U"). When executed, uninitialized nodes are replaced by type-specific nodes, such as for Integer ("I"), Double ("D"), or String ("S"). These nodes are a specialized thread-safe replacement dynamically optimized by Truffle.

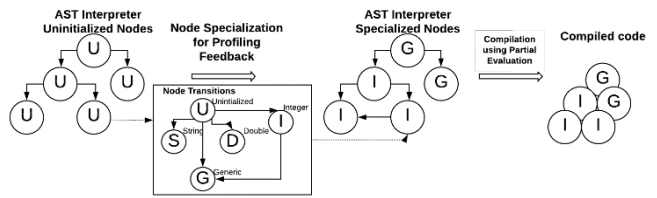

Figure 3. Truffle Optimization Process

Each specialized node has its guard for verification of the validity of the specialization node on each execution. If a guard recognizes node specialization as a failure, then that node will be replaced with a generic version ("G"), which can resolve any possible case, but only a few replacements are allowed for each node. Node Transitions inset could end with completely generic implementation if needed. When nodes in AST are stabilized, Graal compiles the interpreter using partial evaluation and dynamically optimizes it, which produces highly specialized machine code, noted as "Compiled code" [12].

Another possible event is dynamic deoptimization which is triggered by guard failure in optimized machine code. Compiled code is transferred back to AST Interpreter and speculatively optimized more aggressively, i.e., recompiled using partial evaluation.

At runtime, using Truffle, same interoperability protocols are being used by all interpreters. In addition, Truffle virtualizes languages implementation, and this means that from Truffle's perspective there is no significant difference between languages. Therefore, ultimately, all runtime-based tools such as debuggers, profilers, and dynamic analyzers, can be used as polyglot.

*F. Native images and Substrate VM*

GraalVM allows Ahead-Of-Time (AOT) compilation of applications to native images. Substrate VM[8] (SVM) is an embeddable VM with integrated native development tools, written in Java and optimized to execute Truffle compatible languages. It has features any VM would have but precompiled by Graal. Java application is being precompiled together with SVM, resulting in a native image of an entire application. Internally, this is a binary that has VM packaged together with the Java program. Usage of native images can have significant performance benefits in certain situations, when JVM is not required to run, load, and initialize classes, resulting in a fast startup and low footprint. Such executables do not express with peak performance, but a fast startup and low runtime overhead could make a significant difference in a contemporary cloud or serverless production environment.

*G. Embeddable*

GraalVM extends its polyglot capabilities by being able to embed itself into different runtime platforms. Taken as a binary, GraalVM can be put into a Java VM, as standalone or embedded into a database. Possible implementations include MySQL and Oracle databases, Apache Spark, NGINX, and others. For example, in an experimental build of Oracle database, GraalVM runtime was integrated [13], which resulted in much simpler business logic, by enabling users to call a JavaScript function directly from SQL query directly. The advantages of embeddability introduce the ability to push business logic straight into the database by skipping intermediate steps. In addition, programming languages, modules, and libraries of a software ecosystem that user may work with are supported and can be in further development of applications.

### III. THE BENCHMARK AND THE RESULTS

*A. The Benchmark*

The DaCapo benchmark suite [9] is a tool for Java benchmarking mainly created for performance analysis of JVMs, including compiler and memory management analysis. The benchmark suite consists of open source real-world benchmark application in order to showcase an environment similar to production. As a reference guide for comparison we used a study on Java Benchmarks [14], and research on GC performance has been taken into consideration when choosing viable benchmarking tests [15]. We used a set of benchmark tests from DaCapo suite version 9.12-MR1-bach to get a wider spectrum of results.

The benchmark cases used in testing are the following:

- *h2* – execution of JDBC like in-memory benchmark, by running a large number of transactions on a banking application model
- *lusearch-fix* – *an* application that indexes works of Shakespeare and the King James Bible by using *lucene*
- *xalan* – an application that transforms XML documents into HTML documents
- *pmd* – an application that analyzes a number of Java classes for a range of source code problems
- *sunflow* – an application that renders a set of images using ray tracing
- *jython* – an application that interprets the *pybench* Python benchmark

Since DaCapo suite's system automatically detects a steady state where results vary with a much smaller deviation, tests need to be run in a number of iterations *N*, and benchmark takes *N-1* iterations as a warmup, with final iteration being used as the most credible. In the same manner, a certain number of warmups is needed for the JIT compilation and GC mechanism to stabilize.

*B. Testing environment configuration*

Our environment configuration consists of OpenJDK 64-Bit Server VM build 11.0.2+9-Debian-3 mixed mode with same Javac version, OpenJDK 64-Bit Server VM build 10.0.2+13-Debian-1 mixed mode with same Javac version, and for GraalVM we used GraalVM Enterprise Edition (EE), GraalVM 1.0.0-rc11 build 25.192-b12-jvmci-0.53, mixed mode with Open JDK version 1.8.0_192, and GraalVM Community Edition (CE), GraalVM 1.0.0-rc11 build 25.192-b12-jvmci-0.53 mixed mode with Open JDK version 1.8.0_192. OS we were using as test environment is Linux Kali 4.17.0-kali1-amd x86_64 GNU/Linux. We have complied with GraalVM's instructions to install both CE and EE version of GraalVM, to investigate possible differences. Every benchmark application was run by GNOME Terminal version 3.28.2.

TABLE II. TEST RESULTS FOR SELECTED DACAPO BENCHMARKS WITHIN GRAAL EE, GRAAL CE, JDK 10, AND JDK 11 ENVIRONMENTS WITH TIME (IN MILLISECONDS)

| Benchmarks | Graal EE | Graal CE | JDK 10 | JDK 11 |
|---|---|---|---|---|
| *h2 avg* | 4494.95 | 5327.10 | 10821.85 | 9550.05 |
| *h2 stdev* | 613.18 | 336,18 | 618.49 | 515.68 |
| *lusearch avg* | 2683.40 | 2757.10 | 2874.50 | 2886.15 |
| *lusearch stdev* | 223.60 | 378.66 | 264.96 | 203.40 |
| *xalan avg* | 863.32 | 277.41 | 870.72 | 232.51 |
| *xalan stdev* | 863.32 | 277.40 | 870.72 | 232.51 |
| *pmd avg* | 138.23 | 194.61 | 201.95 | 276.76 |
| *pmd stdev* | 138.29 | 194.61 | 201.94 | 276.70 |
| *sunflow avg* | 5062.05 | 5626.25 | 5417.25 | 6151.25 |
| *sunflow stdev* | 630.12 | 536.45 | 274.53 | 307.89 |
| *jython avg* | 6652.86 | 7038.34 | 6433.60 | 6672.10 |
| *jython stdev* | 2325.62 | 2095.47 | 2159.1 | 1667.34 |

---

[8] Substrate VM, https://github.com/oracle/graal/tree/master/substratevm

[9] DaCapo Benchmarks, http://dacapobench.org/

*C. Test cases*

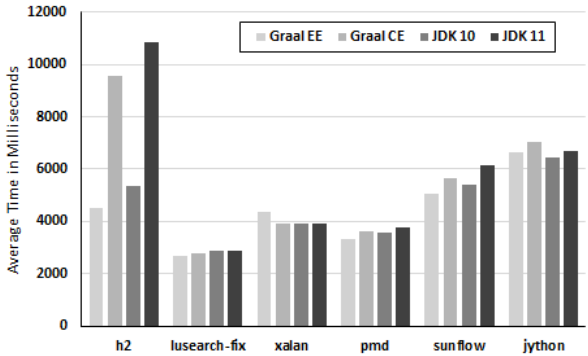

Figure 4. Preliminary average test results for selected DaCapo benchmarks on Graal EE, Graal CE, JDK 10, and JDK 11

Preliminary test results reported in Table II present stable runs with average and standard deviation test time expressed in milliseconds. All benchmarks have been warmed up before taking precise measurement since it takes time for benchmark and JIT compiler to stabilize caches. Therefore, we have mostly recorded the best performance that resulted in our environment, defined by the lowest time in milliseconds it took the test program to run. Each benchmark was performed in 30 iterations, and several initial runs were ignored, and this was repeated 4 times. The test included environments Graal EE, Graal CE, JDK 10, and JDK 11, with their respected version mentioned in the previous section. Due to limited resources in our academic environment, our preliminary results represent a smaller number of iterations tested on limited computing power.

*D. Preliminary Results and Discussion*

Within our test environments, preliminary results describe the following. As presented in Fig. 4, Graal EE performs the best in 4 out of 6 cases. In *xalan* and *jython* benchmark tests are the only ones where Graal EE doesn't have the best results, but it is by a small difference compared to JDK 10 benchmark result. In benchmark *xalan* JDK 10 performs 11.17% better, and in benchmark *jython* 3.4% better. On the other hand in benchmark *h2,* GraalVM is better by as much as 19.8%.

On the other hand, it is interesting how Graal CE behaves in a given environment, mostly similar to JDK 11, but particularly in *h2* benchmark test as they both result with much higher values compared to Graal EE and JDK 10, for example Graal EE completes *h2* benchmark in 47.06% faster than Graal CE. In addition, Graal EE performs the worst in *xalan* application test; however, Graal CE performs similarly to JDK 10 and JDK 11.

Considering standard deviation, Graal CE turned out to be more consistent than Graal EE and JDK 10, but the outcome from JDK 11 presents it as being the most stable. Correspondingly to our premise from the previous section Graal CE and JDK 11 perform in a similar manner. An interesting result is that in *pmd* application test both Graal CE and EE performed significantly better than JDKs. In total, JDK 11 is the most stable version with Graal CE following closely.

As our results present, using GraalVM has many advantages. In 4 out of 6 tests in which GraalVM outperforms the competition we found a similar software architectural pattern, and it is important to understand how this affects our results.

Benchmark tests *h2*, *sunflow*, and *pmd* are all multithreaded applications where our results are similar, with Graal EE being the fastest, JDK 10 and Graal CE following closely, and JDK 11 being the last. In addition, multithreaded application result within *lusearch-fix* and *xalan*, at least one of GraalVM versions is better than standard JDK. On the contrary, *jython* test as a single thread application shows the best result with JDK 10.

In addition to benchmark testing, we have also compared Graal's polyglot performance within our environment, and the preliminary results were successful. Our tests mainly include test examples conducted by the GraalVM team, such as running simple applications in different programming languages to test Graal's interoperability features. A particular example was using Java's BigInteger objects in JavaScript, and JavaScript's regular expression used in Python string matching. We have experimented with these examples in different configurations, and it worked as expected without larger wait times or any errors.

IV. FUTURE WORK

Our initial research was of a simpler scope covering basic functionality and features of GraalVM. Our preliminary test results were conducted in a limited environment thus we plan to continue this research with experiments conducted on a much larger scale. An interesting part of this project are additions in the form of stochastic performance optimizations named Genetic Improvement (GI), which could be further investigated. Additionally, our research in the current state could be extended more to cover the detailed process in creating native images and embedding process.

V. CONCLUSION

With the current version of GraalVM, without any additional configuration and tuning, in comparison to standard JDK, both Graal CE and Graal EE versions proved to be better in most cases. Tests in which GraalVM performed slightly worse as standard JDK can be mostly disregarded since the difference is negligible and, at this point in time, without discovering the cause. This indicates that with the next version of GraalVM we can expect further optimizations and presumably even better results. With GraalVM being recently deployed it means that it works better with modernized syntax and algorithms; thus, using contemporary software practices our performance results could be even better. Considering the size of this project and all additional features it provides we can say that performance wise, GraalVM executes praiseworthy.

GraalVM's polyglot features show interesting promises for future usage and development, furthermore, providing the Cross-Language Interoperability support for polyglot programming in many supported languages. Thus, for example, a project could be written in several programming languages without any performance overhead allowing computing problems to be solved in the best-specialized language.

We believe the most prominent features of GraalVM project are native images and embeddability options. As contemporary software includes the use of databases, cloud and serverless services, these features could become imperative in the future of development.

GraalVM is still in development, but we have promising results with various tests and with features that work as defined, and as such we have optimistic expectation in the next few years of development.